\documentstyle [12pt] {article}

\catcode`@=12
\input epsf.tex
\def\DESepsf(#1 width #2){\epsfxsize=#2 \epsfbox{#1}}

\begin{document}
\thispagestyle{empty}
\begin{flushright} UCRHEP-T199\\August 1997\
\end{flushright}
\vspace{0.5in}
\begin{center}
{\Large	\bf Decay of Z into Three Pseudoscalar Bosons\\}
\vspace{1.5in}
{\bf E. Keith and Ernest Ma\\}
\vspace{0.3in}
{\sl Department of Physics\\}
{\sl University of California\\}
{\sl Riverside, California 92521\\}
\vspace{1.5in}
\end{center}
\begin{abstract}\
We consider the decay of the $Z$ boson into three pseudoscalar bosons in a 
general two-Higgs-doublet model.  Assuming $m_A$ to be very small, and that 
of the two physical neutral scalar bosons $h_1$ and $h_2$, $A$ only couples 
to $Z$ through $h_1$, we find the $Z \rightarrow A A A$ branching fraction 
to be negligible for moderate values of $\tan \beta \equiv v_2/v_1$, 
if there is no $\lambda_5 
(\Phi_1^\dagger \Phi_2)^2 + h.c.$ term in the Higgs potential; otherwise 
there is no absolute bound but very large quartic couplings (beyond the 
validity of perturbation theory) are needed for it to be observable.
\end{abstract}

\newpage
\baselineskip 24pt

If the standard $SU(2) \times U(1)$ electroweak gauge model is extended to 
include two scalar doublets, there will be a neutral pseudoscalar boson $A$ 
whose mass may be small.  In that case, the decay of the $Z$ boson into 
3 $A$'s may not be negligible.  This process was first studied\cite{1} in 
a specific model\cite{2}.  It was then discussed\cite{3} in a more general 
context.  More recently, it has been shown\cite{4} that there is a lower 
bound on $m_A$ of about 60 GeV in the Minimal Supersymmetric Standard Model 
(MSSM), hence the decay $Z \rightarrow AAA$ is only of interest for models 
with two scalar doublets of a more general structure.  Even in the context of 
supersymmetry, this is possible\cite{5} if there exists an additional 
U(1) gauge factor at the TeV scale.  

In this paper we consider a general two-Higgs-doublet model and identify 
the conditions for which the decay $Z \rightarrow AAA$ may be enhanced, 
despite the nonobservation of $e^+ e^- \rightarrow h + A$, where $h$ is 
either one of the two neutral scalar bosons of the model.  We will show 
that in principle this decay is limited only by the scalar coupling 
$\lambda_1 - \lambda_2$ as defined below.  However, if 
$\lambda_5 = 0$, which is true in a large class of models\cite{6}, then 
it may be bounded as discussed below.

Let the Higgs potential $V$ for two $SU(2) \times U(1)$ scalar doublets 
$\Phi_{1,2} = (\phi^+_{1,2}, \phi^0_{1,2})$ be given by
\begin{eqnarray}
V &=& m_1^2 \Phi_1^\dagger \Phi_1 + m_2^2 \Phi_2^\dagger \Phi_2 + m_{12}^2 
(\Phi_1^\dagger \Phi_2 + \Phi_2^\dagger \Phi_1) + {1 \over 2} \lambda_1 
(\Phi_1^\dagger \Phi_1)^2 + {1 \over 2} \lambda_2 (\Phi_2^\dagger \Phi_2)^2 
\nonumber \\ &+& \lambda_3 (\Phi_1^\dagger \Phi_1)(\Phi_2^\dagger \Phi_2) + 
\lambda_4 (\Phi_1^\dagger \Phi_2)(\Phi_2^\dagger \Phi_1) + {1 \over 2} 
\lambda_5 (\Phi_1^\dagger \Phi_2)^2 + {1 \over 2} \lambda_5^* (\Phi_2^\dagger 
\Phi_1)^2,
\end{eqnarray}
where the discrete symmetry $\Phi_1 \rightarrow \Phi_1$ and $\Phi_2 
\rightarrow -\Phi_2$ is only broken softly by the $m_{12}^2$ term.  
Assume $\lambda_5$ to be real for simplicity.  Define $\tan \beta \equiv 
v_2/v_1$ as is customary, where $v_{1,2} = \langle \phi_{1,2}^0 \rangle$ 
are the usual two nonzero vacuum expectation values.  The pseudoscalar 
neutral Higgs boson is then
\begin{equation}
A = \sqrt 2 (\sin \beta {\rm Im} \phi_1^0 - \cos \beta {\rm Im} \phi_2^0),
\end{equation}
with mass given by
\begin{equation}
m_A^2 = -m_{12}^2 (\tan \beta + \cot \beta) - 2 \lambda_5 v^2,
\end{equation}
where $v^2 \equiv v_1^2 + v_2^2$, and the charged Higgs boson is 
\begin{equation}
h^\pm = \sin \beta \phi_1^\pm - \cos \beta \phi_2^\pm,
\end{equation}
with
\begin{equation}
m_{h^\pm}^2 = m_A^2 + (\lambda_5 - \lambda_4) v^2.
\end{equation}
To get the maximum $Z \rightarrow AAA$ rate, we let $m_A = 0$, {\it i.e.}
\begin{equation}
m_{12}^2 = -2 \lambda_5 v^2 \sin \beta \cos \beta.
\end{equation}
Then the mass-squared matrix spanning the two neutral scalar Higgs bosons 
$\sqrt 2 {\rm Re} \phi_{1,2}^0$ is given by
\begin{equation}
{\cal M}^2 = 2 v^2 \left[ \begin{array} {c@{\quad}c} \lambda_1 \cos^2 \beta + 
\lambda_5 \sin^2 \beta & (\lambda_3 + \lambda_4) \sin \beta \cos \beta \\ 
(\lambda_3 + \lambda_4) \sin \beta \cos \beta & \lambda_2 \sin^2 \beta + 
\lambda_5 \cos^2 \beta \end{array} \right].
\end{equation}
Consider now the following two linear combinations:
\begin{eqnarray}
h_1 &=& \sqrt 2 (\sin \beta {\rm Re} \phi_1^0 - \cos \beta {\rm Re} \phi_2^0),
\\ h_2 &=& \sqrt 2 (\cos \beta {\rm Re} \phi_1^0 + \sin \beta {\rm Re} 
\phi_2^0).
\end{eqnarray}
It is well-known that $h_1$ couples to $AZ$ but not $ZZ$, whereas $h_2$ 
couples to $ZZ$ but not $AZ$. However, the process $e^+ e^- \rightarrow 
h + A$ is in general possible because $h$ will normally have a $h_1$ 
component, thereby putting a constraint on $m_A$ if kinematically allowed.  
For our purpose, we will require $h_1$ and $h_2$ 
to be mass eigenstates, in which case $m_A$ is uncontrained by the 
nonobservation of $e^+ e^- \rightarrow h + A$ even if $m_2$ is small, as 
long as $m_1$ is larger than the $e^+ e^-$ center-of-mass energy.  This 
allows us to have the maximum effective coupling of $Z$ to $AAA$ as 
shown below.

The requirement that $h_1$ and $h_2$ be mass eigenstates leads to the 
condition
\begin{equation}
\lambda_2 \sin^2 \beta - \lambda_1 \cos^2 \beta + (\lambda_3 + \lambda_4 + 
\lambda_5) (\cos^2 \beta - \sin^2 \beta) = 0.
\end{equation}
As a result, the masses of $h_{1,2}$ are given by
\begin{eqnarray}
m_1^2 &=& [\lambda_1 \cos^2 \beta + \lambda_2 \sin^2 \beta + \lambda_5 
- \lambda_3 - \lambda_4] v^2,  \\ m_2^2 &=& [\lambda_1 \cos^2 \beta + 
\lambda_2 \sin^2 \beta + \lambda_5 + \lambda_3 + \lambda_4] v^2.
\end{eqnarray}
Note that in the MSSM, Eq.~(10) cannot be satisfied in the presence of 
radiative corrections.

We now extract the $h_1 A A$ coupling from Eq.~(1), using Eqs.~(2) and (8). 
We find it to be given by
\begin{equation}
{{\sin 2 \beta} \over {2 \sqrt 2}} (\lambda_1 - \lambda_2) v,
\end{equation}
where Eq.~(10) has been used. 
As a function of $\beta$, this expression is obviously maximized at 
$\sin 2 \beta = \pm 1$.  On the other hand, our conditions so far do 
not limit the combination $\lambda_1 - \lambda_2$, hence 
there is no absolute bound on $Z \rightarrow AAA$ in this general case.

Let us consider the case $\lambda_5 = 0$.  This is natural in a large 
class of models where the two Higgs doublets are remnants\cite{6} of a gauge 
model larger than the standard model such that they are distinguishable 
under the larger symmetry.  In that case, we have
\begin{equation}
m_1^2 = 2 (\lambda_1 - \lambda_3 - \lambda_4) v^2 \cos^2 \beta = 
2 (\lambda_2 - \lambda_3 - \lambda_4) v^2 \sin^2 \beta,
\end{equation}
and we can rewrite (13) as
\begin{equation}
- {{m_1^2} \over {v \sqrt 2}} \cot 2 \beta.
\end{equation}
The above expression appears to be unbounded as $\sin 2 \beta \rightarrow 0$. 
However, that would require very large quartic scalar couplings.  This can be 
seen two ways.  First, since (15) is equal to (13), we need an extremely large 
value of $\lambda_1 - \lambda_2$.  Second, from Eq.~(14), we see also that if 
$\sin \beta$ is small, then $\lambda_2 - \lambda_3 - \lambda_4$ has to be 
big, and if $\cos \beta$ is small, then $\lambda_1 - \lambda_3 - \lambda_4$ 
has to be big.  Thus we will choose moderate values of $\tan \beta$ in (15) 
for the following discussion.

In Figure 1 we show the diagram for the decay $Z \rightarrow AAA$ with 
an intermediate virtual $h_1$. To maximize this rate, we minimize 
$m_1$ to be just above the maximum experimental $e^+ e^-$ center-of-mass 
energy, which is 172 GeV up to now but will soon be 183 GeV.  As for $h_2$, 
it interacts exactly as the one Higgs boson of the standard-model, from 
which we have the experimental limit\cite{7} of $m_2 > 65$ GeV.  However, 
$m_2$ is not directly involved in the $h_1 A A$ coupling here.  Note also 
that $\lambda_4$ by itself must be large and negative so that $m_{h^\pm}$ 
of Eq.~(5) can be greater than $m_t - m_b$ for $m_A = 0$, so as to prevent 
the decay $t \rightarrow b + h^+$.  This condition is not satisfied in the 
MSSM where $\lambda_4 = -g_2^2/2$, hence $m_A = 0$ is not allowed 
there\cite{4}.

Assuming $\lambda_5 = 0$ and using Eq.~(15) with $m_1 = 180$ GeV and 
$|\cot 2 \beta| = 1$ ({\it i.e.} $\tan \beta = 0.4$ or 2.4), we now 
calculate the $Z \rightarrow AAA$ decay rate, following Ref.~[1].  
The amplitude is given by
\begin{equation}
{\cal M} = g_Z {{m_1^2 \sqrt 2} \over v} \left[ {{\epsilon \cdot 
k_1} \over {(p - k_1)^2 - m_1^2}} + {{\epsilon \cdot k_2} \over {(p - k_2)^2 
- m_1^2}} + {{\epsilon \cdot k_3} \over {(p - k_3)^2 - m_1^2}} \right],
\end{equation}
where $g_Z = e/\sin \theta_W \cos \theta_W$, $p$ is the four-momentum of the 
$Z$ boson, and $k_{1,2,3}$ are those of the $A$'s.  The effective 
coupling used in Ref.~[1] is now determined to be
\begin{equation}
\lambda_{\rm eff} = {{m_1^2 \sqrt 2} \over v^2} \simeq 1.5.
\end{equation}
Using the estimate of Ref.~[1], 
this $Z \rightarrow AAA$ rate is then about $1.0 \times 10^{-7}$ GeV. 
Hence its branching fraction is about $4 \times 10^{-8}$ which is 
clearly negligible.  To obtain a branching fraction of $10^{-6}$, we need 
$\cot 2 \beta = 5$ ({\it i.e.} $\tan \beta = 0.1$ or 10).  In this case, 
either $\lambda_1 - \lambda_3 - \lambda_4$ or $\lambda_2 - \lambda_3 - 
\lambda_4$ in Eq.~(14) has to be about 53.5. 
If $\lambda_5 \neq 0$, then we cannot use Eqs.~(14) and (15), but Eq.~(13) 
is still valid.  To obtain a branching fraction of $10^{-6}$, we will then 
need $|\lambda_1 - \lambda_2|$ to be about 53.5.  Thus 
in both scenarios, one or more quartic scalar couplings have to be very 
large and beyond the validity of perturbation theory.

If $h_1$ and $h_2$ are not exact mass eigenstates, then there is an additional 
contribution from $h_1 - h_2$ mixing which is necessarily very small from the 
constraint of experimental data if $m_2$ is below 172 GeV.  The $h_2 A A$ 
coupling is given by
\begin{equation}
{v \over \sqrt 2} \left( {m_2^2 \over {2 v^2}} - 2 \lambda_5 [1 - \sin^2 \beta 
\cos^2 \beta] \right).
\end{equation}
If $\lambda_5 = 0$, this expression is bounded independent of $\tan \beta$ 
and the overall contribution (including the small $h_1 - h_2$ mixing) is 
negligible.  If $\lambda_5 \neq 0$, then its value has to be huge 
for the process to be observable.

The reason that $\Gamma (Z \rightarrow AAA)$ is so small is twofold.  One 
is that with the higher energy reached by LEP2, the nonobervation of 
$Z \rightarrow h + A$ forces $m_1$ to be much greater than $M_Z$.  The 
other is that for $m_1 >> M_Z$, the leading term in $\cal M$ vanishes 
because $\epsilon \cdot (k_1 + k_2 + k_3) = 0$, resulting in a very 
severe suppression factor\cite{1}.  Our conclusion is 
that the decay $Z \rightarrow AAA$ is not likely to be observable 
in a general two-Higgs-doublet model with parameters in the perturbative 
regime.
\vspace{0.5in}

\begin{center} {ACKNOWLEDGEMENT}
\end{center}

This work was supported in part by the U. S. Department of Energy under 
Grant No. DE-FG03-94ER40837.

\newpage
\bibliographystyle{unsrt}

\vspace{0.5in}
\begin{center} {\bf Figure Caption}
\end{center}

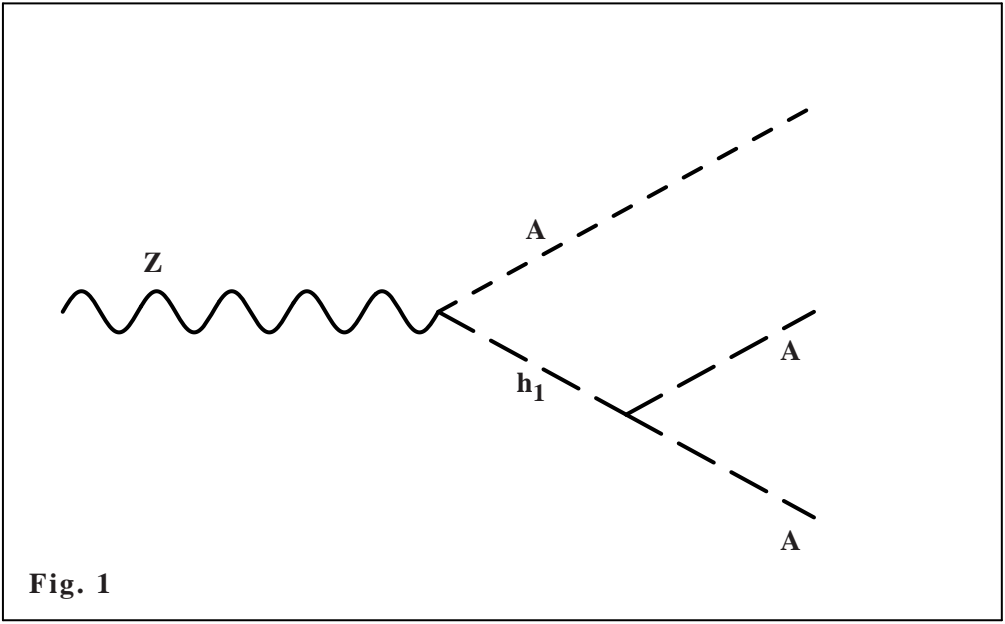
\noindent Fig.~1.  One of 3 diagrams for the decay $Z \rightarrow AAA$.  The 
other 2 are obvious permutations.

\newpage
\begin{figure}[htb]
\centerline{ \DESepsf(aaa.epsf width 15 cm) } \smallskip
\nonumber
\end{figure}

\end{document}